\documentclass[twocolumn,showpacs,preprintnumbers,prb,aps]{revtex4}
\usepackage{graphicx}
\usepackage{dcolumn}
\usepackage{amsfonts,amsmath,amssymb,bm}
\usepackage{natbib}
\usepackage{hyperref}
\setcitestyle{super}

\begin{document}

\title{\bf {Highly Anistoropic Thermal Conductivity of Arsenene: An \textit{ ab initio} Study}}
\date{\today} 
\author{M.~Zeraati$^a$}
\author{S.~M.~Vaez Allaei$^{a,b}$}
\author{I. Abdolhosseini Sarsari$^c$}
\author{M.~Pourfath$^{d,e}$}
\author{D.~Donadio$^{f,g,h,i}$}

\affiliation{a) Department of Physics, University of Tehran, Tehran 14395-547, Iran\\
b)School of Physics, Institute for Research in Fundamental Sciences (IPM), Tehran 19395-5531, Iran\\
c)Department of Physics, Isfahan University of Technology, Isfahan, 84156-83111, Iran\\
d)School of Electrical and Computer Engineering, University of Tehran, Tehran 14395-515, Iran\\
e)Institute for Microelectronics, TU Wien,  Gu{\ss}hausstra{\ss}e 27--29/E360, 1040 Vienna, Austria\\
f)Department of Chemistry, University of California Davis, One Shields Ave. Davis, CA, 95616\\
g)Max Planck Institut f\"ur Polymerforschung, Ackermannweg 10, D-55128 Mainz, Germany\\
h)Donostia International Physics Center, Paseo Manuel de Lardizabal, 4, 20018 Donostia-San Sebastian, Spain\\
i)IKERBASQUE, Basque Foundation for Science, E-48011 Bilbao, Spain}

\newcommand{\Eq} [1] {Eq.~\ref{#1}}
\newcommand{\Fig}[1] {Fig.~\ref{#1}}

\begin{abstract}
Elemental 2D materials exhibit intriguing heat transport and phononic properties.
Here we have investigated the lattice thermal conductivity of newly proposed arsenene, the 2D 
honeycomb structure of arsenic, using {\it ab initio} calculations.  
Solving the Boltzmann transport equation for phonons, we predict a highly anisotropic thermal
conductivity, of $30.4$ and $7.8$ W/mK along the zigzag and armchair directions, respectively at room temperature.
Our calculations reveal that phonons with mean free paths between $20$ nm and $1$ $\mu$m provide the
main contribution to the large thermal conductivity in the zig-zag direction, mean free paths of
phonons contributing to heat transport in the armchair directions range between $20$ and $100$ nm. 
The obtained low and anisotropic thermal conductivity, and feasibility of synthesis, in addition to other reports
on high electron mobility, make arsenene a promising material for a variety of applications, including
thermal management and thermoelectric devices.
\end{abstract}

\maketitle
The discovery of graphene as a stable atomically thin material has led to extensive investigation of similar 2D systems. 
Its properties such as high electron mobility~\cite{Novoselov2005}, and very high thermal conductivity~\cite{xu2014length,Fugallo2014,Lindsay:2014cg,Cepellotti:2015ke}
make graphene very appealing for applications in electronics, for packaging and thermal
management~\cite{Wei2009,Zuev2009,Wang2011,Rajabpour2011,Rajabpour2012,Yan:2012jg}.
The successful isolation of single-layer graphene fostered the search for further ultra-thin 2D structures, such as silicene,
germanene, phosphorene, and transition metal dichalcogenides, e.g. MoS$_2$ and WS$_2$~\cite{Geim2013,Guan2014}. These materials are now
considered for various practical usages due to their distinguished properties stemming from their low dimensionality. 
Thermal transport in 2D materials has recently attracted the attention of the scientific community, as
anomalous heat conduction has been predicted to occur in systems with reduced dimensionality \cite{Lepri:2003fc}.  
Phononic properties and thermal conductivity vary significantly from one 2D system to
another~\cite{Lindsay2010Flexural,Xie2014Thermal,Li2013Thermal,Jain2015Strongly}. 
For example, silicene has a buckled structure and a lower thermal
conductivity~\cite{Xie2014,Yang2014} compared to graphene~\cite{chen2012evidence,Vogt2012,Geim2013}. 

\begin{figure}[!htb]
	\centerline{\includegraphics[width=0.9\linewidth]{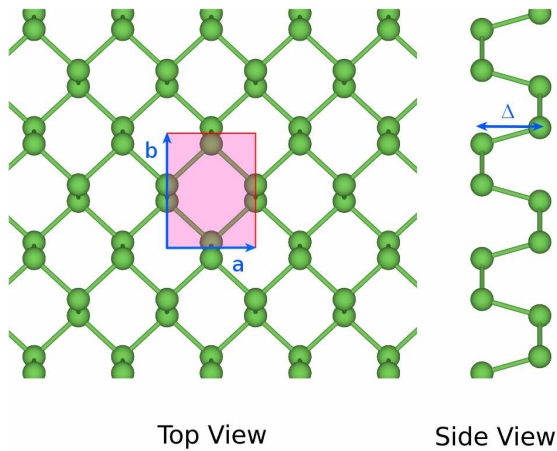}}
	\caption{Top and side view of puckered arsenene.}
	\label{FIG1}
\end{figure}

2D structures of arsenic and phosphorous have been recently investigated~\cite{Kamal2015,Zhang2015,Zhang2015orientation,Zhang2014,Zhu2014Unusual}. 
Arsenic and phosphorus are in the 5th group of the periodic table and both have different allotropes. 
Black phosphorus is a layered allotrope of phosphorus similar to graphite, and the stability of its single layer form, named {\it phosphorene}
has been probed both theoretically and experimentally~\cite{Guan2014,Liu2014}.  
{\it Gray arsenic} is one of the most stable allotropes of arsenic with a buckled layered structure~\cite{Zhu2014Unusual,Madelung2004}.
In addition, arsenic has an orthorhombic phase (puckered) similar to black phosphorus~\cite{Kamal2015,Zhang2015orientation,Zhang2014},
and its monolayer is called {\it arsenene} (see \Fig{FIG1}). Experimental observations have shown that gray arsenic undergoes a structural
phase transition to the orthorhombic precursor of arsenene at temperatures of about $T=370$ K~\cite{Krebs1957}. 
As a monolayer arsenene has a direct band gap as opposed to the multilayer allotrope, which exhibits an indirect band gap of the order of
$1$~eV~\cite{Zhang2014}. According to our calculations, monolayer of arsenene is stable also near zero temperature, in agreement with
previous reports~\cite{Kamal2015,Zhang2015orientation,Zhang2014}. 
Here we investigate heat conduction in arsenene and we elucidate the anisotropy of its thermal conductivity. Since in semiconductors phonons are the predominant heat carriers, we use first-principles anharmonic lattice dynamics calculations and the Boltzmann transport equation to compute phonon dispersion relations and thermal conductivity.

We employ density functional theory as implemented in the VASP code~\cite{kresse1996} with projector-augmented-wave 
(PAW)~\cite{Blochl1994} pseudopotentials and the Perdew-Burke-Ernzerhof (PBE) generalized gradient functional
~\cite{perdew1996generalized} exchange-correlation functional. Integration over Brillouin zone is performed
using $15\times15\times1$ $\Gamma$-center Monkhorst-Pack mesh of k-points~\cite{monkhorst1976special} and plane waves up to an energy cutoff of $360$ eV are used as basis set to represent the Kohn-Sham wave functions. The simulation cell is built with a vacuum layer of
$10$ \AA{}, which is sufficient to avoid interactions between periodic images. 
The structural relaxations were performed by using the conjugate-gradient (CG) algorithm with a convergence criterion of
$10^{-6}$ eV/\AA{} for the maximum residual force component per atom.

\begin{figure}
	\includegraphics [width=0.9\linewidth]{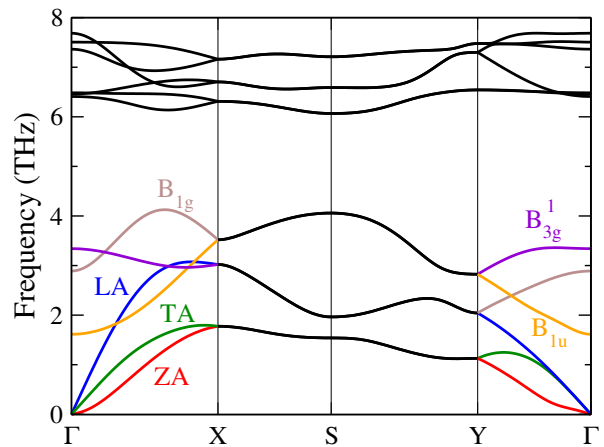}
	\caption{(a) Phonon dispersion curves along the high-symmetry directions of the first Brillouin zone for puckered arsenene.}
	\label{FigDispersion}
\end{figure}

As shown in \Fig{FIG1}, arsenene has a puckered structure arranged in a rectangular unit cell with lattice
vectors $a=3.686$ \AA{} and $b=4.769$ \AA{}. The bond length between atoms in the same plane (upper panel or lower)
is $2.510$ \AA{}, while the bonding length between atoms in different planes is $2.496$ \AA{}. The distance between
the two planes is denoted by $\Delta$ and is $2.402$ \AA{}. Our structural parameters are compared with previous
studies in Table \ref{t:lattice}. 
\begin{table}
\caption{\label{t:lattice} The lattice parameters of arsenene. $a$ and $b$ are the lattice vectors, $\Delta$ is
the puckering distance, $c_1$ is the bond length between atoms that are in the same plane, $c_2$ is the bond length
between atoms that are in different planes.}
\begin{tabular}{l|ccccc}
\hline\hline
\multicolumn{6}{c}{Mono-layer of arsenene}\\ \hline
 		Method                                       & $\enspace$ $a$ (\AA) & $b$ (\AA) & $\Delta$ (\AA) & $c_1$ (\AA) & $c_2$ (\AA) \\ \hline
 		PBE                                          &  $\enspace$ $3.686$  &  $4.769$  &    $2.402$     &   $2.510$   &   $2.496$   \\
 		PBE (Ref.~\cite{Kamal2015})                  &  $\enspace$ $3.677$  &  $4.765$  &       --       &   $2.521$   &   $2.485$   \\
 		revPBE-vdW (Ref.~\cite{Zhang2014})$\enspace$ &  $\enspace$ $3.68$   &  $4.80$   &       --       &   $2.51$    &   $2.49$    \\ \hline\hline
 	\end{tabular}
\end{table}
We have utilized the PHONOPY package~\cite{togo2008first} interfaced with VASP to compute inter-atomic force constants (IFCs).
After testing convergence of the phonon frequencies as a function of the size of the supercell, a $8 \times 8 \times 1$ super-cell has been adopted,  with a $ 3\times3\times1$ $\Gamma$-center Monkhorst-Pack mesh of k-points, to build the dynamical matrix $\mathcal{D}$. 
The frequencies of the vibrational modes are obtained
by diagonalizing $\mathcal{D}$, and the  phonon dispersion curves are represented in \Fig{FigDispersion}. The absence of phonon
branches with negative phonon frequencies indicates the stability of the system at zero temperature.


The phonon bands exhibit two acoustic modes, a longitudinal one (LA) and an in-plane transverse one (TA), with linear dispersion
($\omega\propto q$) for $q~\rightarrow 0$, and an out-of-plane flexure mode (ZA) with quadratic dispersion in the long wavelength
limit, which is a general feature of 2D membranes~\cite{Zabel2001} and have been characterized in
graphene~\cite{Jiang2015A,Lindsay2010Flexural},  silicene~\cite{cahangirov2009two,Xie2014Thermal}, hBN~\cite{Lindsay2011Enhanced},
MoS$_2$~\cite{Cai2014} and ultra thin silicon membranes \cite{Neogi:2015cp,Neogi:2015vj}.  
Flexural modes have a major role in contributing to the thermal conductivity of graphene, both as carriers~\cite{Lindsay2010Flexural}
and as scatterers~\cite{Pereira2013}. In addition, we observe a gap of about $2$ THz between the low-frequency and high-frequency
optical branches. 

\begin{figure}
	\includegraphics[width=0.9\linewidth]{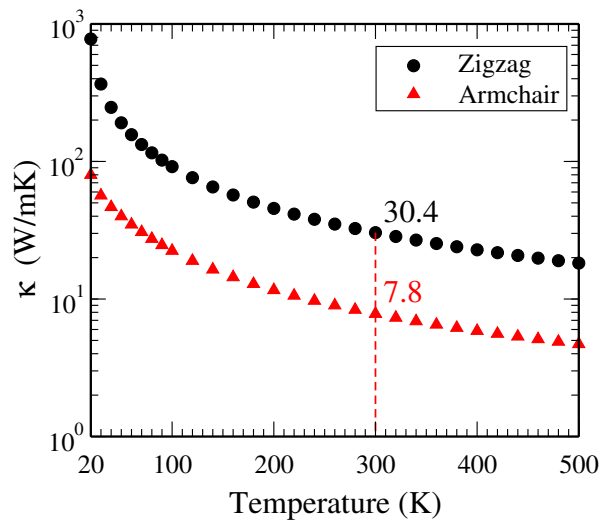}
	\caption{The thermal conductivity of arsenene as a function of temperature along the zigzag and armchair direction.}
	\label{FigKappa} 
\end{figure}
We have computed the thermal conductivity of arsenene by  solving the phonon Boltzmann transport equation
(BTE) with an iterative self-consistent algorithm~\cite{peierls1929kinetic,Omini1996}, using the ShengBTE package~\cite{Li2014ShengBTE}. 
In our calculations we have
considered third-order interatomic force constants up to the forth shell of neighbors in a $5\times5\times1$ super-cell
with a $3\times 3\times1$, $\Gamma$-centered Monkhorst-Pack mesh of k-points~\cite{Li2014ShengBTE,Esfarjani2008}.
Intrinsic three phonon scattering processes have been considered in the calculation of phonon relaxation times:  
\begin{equation}
\frac{1}{\tau^0_\lambda}=\frac{1}{N}
\left(\sum_{\lambda^\prime\lambda^{\prime\prime}}^{+}\Gamma^+_{\lambda\lambda^\prime\lambda^{\prime\prime}}
+\frac{1}{2}\sum_{\lambda^\prime\lambda^{\prime\prime}}^{-}\Gamma^-_{\lambda\lambda^\prime\lambda^{\prime\prime}}
+\sum_{\lambda^\prime}\Gamma^\mathrm{ext}_{\lambda\lambda^\prime}
\right) \ ,
\label{relaxation_time}
\end{equation}
where $\Gamma^+_{\lambda\lambda^\prime\lambda^{\prime\prime}}$ and $\Gamma^-_{\lambda\lambda^\prime\lambda^{\prime\prime}}$  represent
the scattering rates due to absorbing and emitting three-phonon processes, respectively. The extrinsic scattering term
$\Gamma^\mathrm{ext}_{\lambda\lambda^\prime}$ is zero in our calculations, since arsenic does not present isotopic disorder,
and neither defect or boundary scattering processes were considered. 
Knowing the distribution function, the phononic thermal conductivity tensor ($\kappa$) is expressed as:
\begin{equation}
\kappa_{\alpha \beta}=\frac{1}{V}\sum_{\lambda}\hbar\omega_\lambda
\frac{\partial f}{\partial T}v^\alpha_\lambda v^\beta_\lambda \tau_\lambda \ ,
\label{Eq.kappa}
\end{equation}
where $f$ is the phonon distribution function,  $\omega_\lambda$ is the phonon angular frequency, $v^\alpha_\lambda$ is
the velocity component along the $\alpha$-direction, and $\tau_\lambda$ is the phonon relaxation time. 

The two diagonal components of the thermal conductivity at temperatures between $20$ and $500$ K are reported in \Fig{FigKappa}.
The thermal conductivity along the zigzag and armchair direction at $T=300$ K is $30.4$ and $7.8$ Wm$^{-1}$K$^{-1}$, respectively.
The nearly 3-fold ratio between the two components indicates a high anisotropy. The thermal conductivity of
aresenene in comparison to graphene and hexagonal boron nitride is very low: for example both experimental and theoretical studies indicate the intrinsic thermal conductivity of graphene is larger than $3000$ W/mK ~\cite{xu2014length,Fugallo2014}. 
Due to the larger atomic weight of As, $\kappa$ of arsenene is also lower than that of phosphorene, which is also strongly anisotropic, theoretically predicted as $110$ and $36$ Wm$^{-1}$K$^{-1}$ in the zigzag and armchair directions, respectively~\cite{Jain2015Strongly}.
\begin{table}
	\caption{\label{t:sound}Sound velocity (group velocity) of acoustic modes of arsenene close to the zone center ($\Gamma$-point). 
	(*) Due to quadratic form of ZA mode near $\Gamma$ point, the magnitude has been averaged over a small reigon interval.}
	\begin{tabular}{c|ccc|c}
		\hline\hline
		                                    \multicolumn{5}{c}{Sound velocity km/s}                                      \\ \hline
		         &                     \multicolumn{3}{c|}{Arsenene}                      &         Phosphorene          \\ \hline
		         &     ZA (*)     &             TA             &              LA              &  LA (ref.~\cite{Jain2015})   \\ \hline
		Armchair & $0.72$ & $\enspace$$2.41$$\enspace$ & $\enspace$ $2.44$ $\enspace$ & $\enspace$ $4.17$ $\enspace$ \\
		 Zigzag  & $0.34$ &           $2.41$           &            $4.77$            &            $7.73$            \\ \hline\hline
	\end{tabular}
\end{table}
%
We note indeed that the group velocity of the acoustic modes of arsenene along both zigzag and armchair directions near the zone center are
about twice as smaller as those of phosphorene (see Table \ref{t:sound}). Since according to Equation~\ref{Eq.kappa} group velocities contribute to $\kappa$ quadratically, we can argue that the smaller $v_\lambda$ is responsible for the difference between the thermal conductivities of arsenene and phosphorene.
\begin{figure}
	\includegraphics[width=\linewidth]{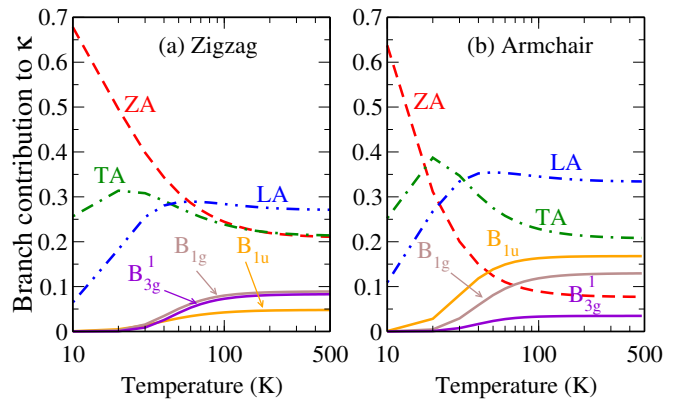}
	\caption{Normalized contribution of each phonon branch to the total thermal conductivity, as a function of temperature along the (a) zigzag and (b) armchair directions.}
	\label{FigContribution} 
\end{figure}

The contribution of each phonon mode to the total thermal conductivity as a function of temperature is shown in Fig. \ref{FigContribution}.
The ZA mode provides the largest contribution to thermal conductivity along both  directions at low temperatures. However, this
contribution decreases as the temperature increases and reaches $7.8$\% (armchair direction) for temperatures above $300$ K. 
This is in contrast to the case of graphene where the ZA modes provide the main contribution to heat conduction~\cite{Lindsay2010Flexural}.
Due to the special selection rules of phonon scattering in graphene, the ZA mode has a relatively large relaxation time
compared to other phonon modes~\cite{Lindsay2010Flexural}, while due to the puckered structure of arsenene, similar to silicene and 
phosphorene~\cite{Xie2014Thermal,Jain2015Strongly}, the hexagonal symmetry of graphene is broken, resulting in larger scattering rates and shorter relaxation times of  the ZA modes at high temperatures. The contribution of the ZA modes remains however substantial and roughly equivalent to that of the TA modes in the zigzag direction, while it drops below 10$\%$ in the armchair direction.

\begin{figure}
	\includegraphics[width=0.8\linewidth]{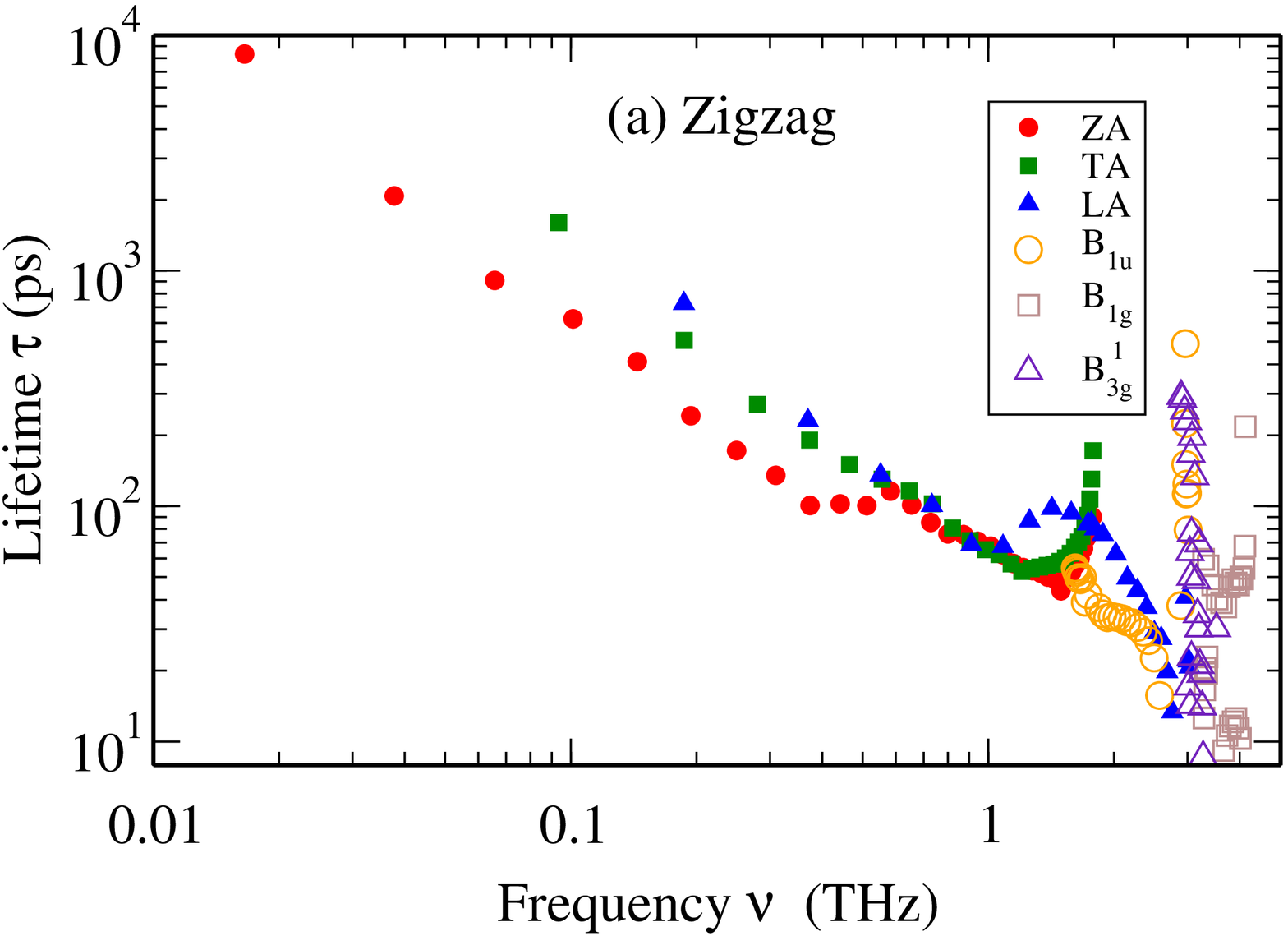}
	\includegraphics[width=0.8\linewidth]{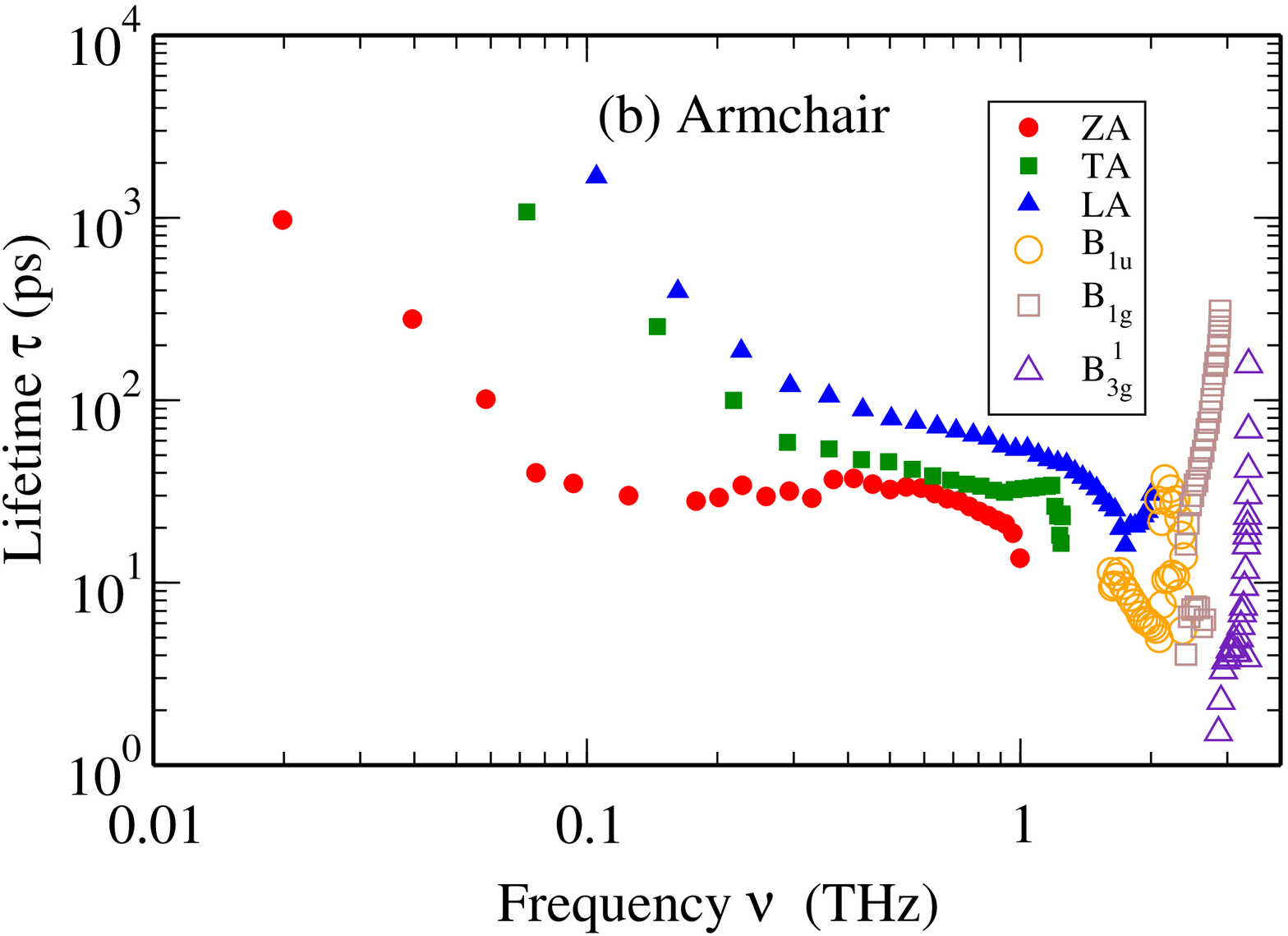}
	\caption{Total relaxation time of phonon modes as a function of frequency $\nu$ along the (a) zigzag  and (b) armchair direction.}
	\label{Figscatt} 
\end{figure}

At temperatures higher than $100$ K, the contribution of LA modes to the thermal conductivity is the largest in both directions. The group velocity of
LA mode near the zone center is the largest along the zigzag direction, while it is comparable to the TA mode along the armchair
direction (see Table \ref{t:sound}). Furthermore, the relaxation times of the LA modes at low frequencies, which play a significant
role in the process of heat conduction, are larger than that of other acoustic modes (see \Fig{Figscatt}). All these effects combined together
result in a larger contribution of LA modes to the thermal conductivity at high temperatures.

\begin{figure}
	\includegraphics[width=0.9\linewidth]{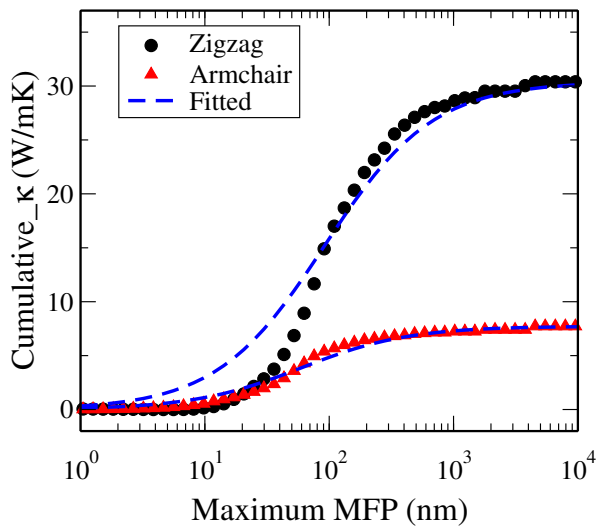} 
	\caption{Cumulative thermal conductivity of arsenene as a function of the phonon mean free path at $T=300$ K along the zigzag and armchair directions. }
	\label{FigKappaCum} 
\end{figure}

Finally we computed the cumulative thermal conductivity as a function of phonon mean free path, defined as
\begin{equation}
\kappa_l (\Lambda \leqslant \Lambda_\mathrm{max})
=\frac{\kappa_l}{1+\frac{\Lambda_0}{\Lambda_{max}}} ,
\label{cumulative}
\end{equation}
where $\Lambda_\mathrm{max}$ is the maximum mean free path and $\Lambda_0$ is a fitting parameter that can be interpreted
as the mean free path of heat-carrying phonons of infinite system size.
By fitting Eq.(\ref{cumulative}) to the calculated values one can evaluate $\Lambda_0$, see \Fig{FigKappaCum}. 
This quantity shows which mean free paths contribute the most to the thermal conductivity. Phonon modes with mean free paths larger
than $59.5$ nm along the armchair direction and $91.8$ nm along the zigzag direction have 50\% contribution to the total thermal conductivity.
Hence we can argue that in a polycrystalline structure with a grain size of the order of $50$ nm the thermal conductivity can be significantly further reduced. However, given the broad range of mean free paths that contribute, the most efficient approach to abate $\kappa$ would be hierarchical nanostructuring~\cite{Biswas:2012fw}.
 
In summary we have theoretically investigated the phonon properties and the  thermal conductivity of
arsenene by {\sl ab initio} anharmonic lattice dynamics. Our calculations evidence the lowest thermal
conductivity among the 2D elemental materials found so far.
The larger mass of arsenic atoms compared to phosphorus leads to a  3 times smaller thermal conductivity
of this material compared to phosphorene.
Similar to phosphorene, the thermal conductivity of arsene is strongly anisotropic. Such anisotropy
could be utilized in thermal engineering application, to establish directional heat transport at the nanoscale.
Unlike graphene, the longitudinal acoustic phonon modes provide the main contribution to the thermal conductivity
of arsenene at temperatures above $100$ K, which is mainly due to the puckered structure of this material. From
this observation, we expect that $\kappa$ of arsenene would be less affected by the interaction with a substrate
than in those materials for which the main contribution comes from flexural modes.
 In addition, given the high electron mobility and the reduced dimensionality, which should enhance the Seebeck
 coefficient, its relatively low and strongly directional thermal conductivity makes arsenene a good candidate from
 thermoelectric energy conversion.  

The work of S.M.V.A. was supported in part by the Research Council of
the University of Tehran.

\bibliography{Paper09arxiveN.bib}
\end{document}